# ERA: Efficient Serial and Parallel Suffix Tree Construction for Very Long Strings

Essam Mansour[1]  Amin Allam[1]  Spiros Skiadopoulos[2]  Panos Kalnis[1]

[1]Math. & Computer Sciences and Engineering
King Abdullah Univ. of Science and Technology
{fname.lname}@kaust.edu.sa

[2]Dept. of Computer Science and Technology
University of Peloponnese
spiros@uop.gr

## ABSTRACT

The suffix tree is a data structure for indexing strings. It is used in a variety of applications such as bioinformatics, time series analysis, clustering, text editing and data compression. However, when the string and the resulting suffix tree are too large to fit into the main memory, most existing construction algorithms become very inefficient.

This paper presents a disk-based suffix tree construction method, called Elastic Range (ERA), which works efficiently with very long strings that are much larger than the available memory. ERA partitions the tree construction process horizontally and vertically and minimizes I/Os by dynamically adjusting the horizontal partitions independently for each vertical partition, based on the evolving shape of the tree and the available memory. Where appropriate, ERA also groups vertical partitions together to amortize the I/O cost. We developed a serial version; a parallel version for shared-memory and shared-disk multi-core systems; and a parallel version for shared-nothing architectures. ERA indexes the entire human genome in 19 minutes on an ordinary desktop computer. For comparison, the fastest existing method needs 15 minutes using 1024 CPUs on an IBM BlueGene supercomputer.

## 1. INTRODUCTION

The suffix tree [12] is a trie that indexes all possible suffixes of a string $S$ (see Figure 1 for an example). It is used to accelerate many string operations. For instance, finding a substring $P$ inside $S$ without an index takes $\mathcal{O}(|S|+|P|)$ time [3]. With a suffix tree the same operation is done in $\mathcal{O}(|P|)$ time, which is a significant gain given that typically $S$ is several orders of magnitude longer than $P$. Other operations that can benefit from a suffix tree include approximate string matching, finding the longest common substring of two strings and finding all common substrings in a database of strings. Such queries are essential for many applications such as bioinformatics [8], time series analysis [15], document clustering [4], text editing [1] and compression [5].

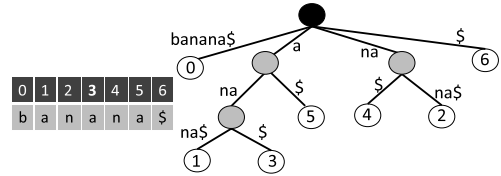

**Figure 1: Suffix tree for $S = banana\$$ ($\$$ denotes end-of-string). Edge labels on a path from the root to a leaf correspond to a suffix in $S$. Leaf labels show the offset of each suffix in $S$**

Fast suffix tree construction is critical, given the rate of data generation by the aforementioned applications [9]. For example, modern DNA sequencers can process multiple samples per hour, whereas financial applications generate continuous streams of time series data. If the string $S$ and the resulting suffix tree can fit in the main memory, there are efficient solutions such as Ukkonen's algorithm [19], which constructs the tree in $\mathcal{O}(|S|)$ time but becomes very inefficient if it runs out of memory. However, the suffix tree for $S$ is more than an order of magnitude larger than $S$. The human genome, for instance, has roughly 2.6G symbols; the resulting suffix tree occupies around 67GB and challenges the main memory limits of many systems. Other algorithms, such as TRELLIS [13], work well if at least $S$ fits in main memory, but are very inefficient otherwise.

In practice $S$ alone can be much larger than the main memory. For example, operations that involve a database of strings [8] require a *generalized* suffix tree, which is simply the suffix tree of the concatenation of all input strings. For such cases, recently two methods have been proposed: $B^2ST$ [2] and WaveFront [7]. Both access $S$ in sequential order which is much faster than random I/Os in modern disks. The serial version of WaveFront is slower than $B^2ST$, but WaveFront is easily parallelizable; this is very important given the size of the targeted problems. Nevertheless, the performance of both algorithms deteriorates as the length of $S$ or the size of the *alphabet* (i.e., set of symbols appearing in $S$) increase.

In this paper we present ERA[1], a suffix tree construction algorithm that *(a)* supports very long strings and large alphabets; *(b)* is much faster than the existing ones even if memory is very limited; and *(c)* is easily parallelizable. In a nutshell, ERA optimizes dynamically the use of memory and amortizes the I/O cost. Specifically, it divides the problem

---
[1]ERA stands for *Elastic Range*, for adjusting dynamically the range of the horizontal partitions.

*vertically* into construction of independent sub-trees, ensuring that each sub-tree can fit into the available memory. Sub-trees are further divided *horizontally* into partitions, such that each partition can be processed in memory with a single sequential scan of the input string $S$. At each step, horizontal partitions are readjusted based on the evolving shape of the tree, in order to maximize memory utilization. Also, vertical partitions may be grouped together in order to share the I/O cost. The entire plan can be executed in a serial or parallel system.

Our contributions include:

- A serial version of ERA that is *at least* 50% faster than existing serial algorithms. Performance gain is more dramatic for very long strings and large alphabets.

- A parallel version for shared-memory and shared-disk architectures that include ordinary multicore desktop systems.

- A parallel version for shared-nothing systems, such as clusters or cloud computing infrastructures.

- Extensive experimental evaluation with real datasets of very long strings. ERA indexes the entire human genome in 19 minutes on an ordinary 8-core desktop computer with 16GB of RAM. For comparison, the fastest existing method (i.e., the parallel version of WaveFront [6]) needs 15 minutes on an IBM BlueGene/L supercomputer using 1024 CPUs and 512GB of RAM.

The rest of this paper is organized as follows. Section 2 furnishes the preliminaries of suffix trees, whereas Section 3 discusses the related work. Section 4 introduces the serial version of ERA and Section 5 presents the parallel versions. Section 6 discusses our experimental results. Section 7 concludes the paper.

## 2. BACKGROUND: SUFFIX TREE

Let $\Sigma$ denote an *alphabet* (i.e., set of symbols). An input string $S$ of length $n+1$ is a sequence $S = s_0 s_1 \cdots s_{n-1}\$$, where $s_i \in \Sigma$, $0 \leq i \leq n-1$ and $\$ \notin \Sigma$; $\$$ is the end-of-string symbol. A *prefix* of $S$ is sequence $s_0 \cdots s_i$ and a *suffix* of $S$, denoted by $S_i$, is $s_i \cdots \$$ ($0 \leq i \leq n$). In this paper, we will consider prefixes of $S$ and prefixes of suffixes of $S$. To avoid confusion, we will refer to the latter by *S-prefixes*. The unique terminal symbol $\$$ ensures that no suffix $S_i$ is a proper S-prefix of any other suffix $S_j$ ($i \neq j$).

A *suffix tree* $\mathcal{T}$ is a trie that indexes all suffixes of $S$. In the rest of the paper, we will use the example string and corresponding tree of Figure 2; the alphabet consists of four symbols $\{A, C, G, T\}$ which is typical in bioinformatics. The main properties of the suffix tree are:

- There exist exactly $n+1$ leaves with node labels from 0 to $n$. For any leaf $v_i$, the concatenation of the edge labels on the path from the root to $v_i$ spells out suffix $S_i$. For example $v_{20}$ corresponds to $S_{20} = TGC\$$.

- Each internal node other than the root, has at least two children and each edge is labeled with a S-prefix of $S$. If, during construction, a node appears with only one child, then the node and its child are merged and the edge labels are concatenated (this explains edge labels with more than one symbol in the example).

- No two edges out of a node can have edge labels beginning with the same symbol.

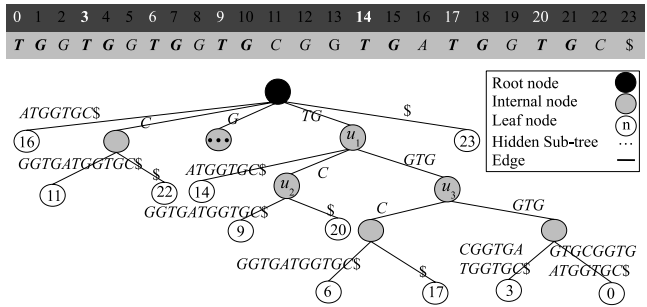

**Figure 2: Input string $S$, where $\Sigma = \{A, C, G, T\}$, and corresponding suffix tree $\mathcal{T}$. For simplicity, $\mathcal{T}_G$ (i.e., the sub-tree under $G$) is not shown**

| $i$ | $S_i$ | **Suffix** |
|---|---|---|
| 0 | $S_0$ | **TG**$GTGGTGGTGCGGTGATGGTGC\$$ |
| 3 | $S_3$ | **TG**$GTGGTGCGGTGATGGTGC\$$ |
| 6 | $S_6$ | **TG**$GTGCGGTGATGGTGC\$$ |
| 9 | $S_9$ | **TG**$CGGTGATGGTGC\$$ |
| 14 | $S_{14}$ | **TG**$ATGGTGC\$$ |
| 17 | $S_{17}$ | **TG**$GTGC\$$ |
| 20 | $S_{20}$ | **TG**$C\$$ |

**Table 1: Suffixes sharing the S-prefix $TG$. $i$ refers to the offset of the suffix in the string of Figure 2**

The suffix tree can be divided into a set of sub-trees; $\mathcal{T}_p$ denotes the sub-tree that indexes suffixes sharing a S-prefix $p$. In the example, $\mathcal{T}$ is divided into $\mathcal{T}_A$, $\mathcal{T}_C$, $\mathcal{T}_G$, $\mathcal{T}_{TG}$, and $\mathcal{T}_\$$. Table 1 shows all suffixes with S-prefix $TG$; these suffixes will be indexed in $\mathcal{T}_{TG}$. The frequency $f_p$ of a S-prefix $p$ is the number of suffixes in $\mathcal{T}_p$. For example, $f_{TG} = 7$, whereas $f_A = 1$. As we will see later, the frequency is proportional to the amount of memory needed for the construction of the sub-tree. Given the available memory, we can bound the maximum frequency of all $p$ below a threshold by using variable length S-prefixes [7]. For example, each of the S-prefixes in the set $\{A, C, TGA, TGC, TGGTGC, TGGTGGTG\}$ has frequency at most 2. Note that, reducing the maximum frequency increases the number of sub-trees.

Storing S-prefixes in the edge labels requires $\mathcal{O}(n^2)$ space for the tree. Typically, a suffix tree edge stores only two integers representing the starting and the ending index of the S-prefix in $S$. Using this representation space complexity drops to $\mathcal{O}(n)$. The figures throughout the paper show S-prefixes for clarity. Also, we sort the edges that emanate from a node according to the lexicographical order of their labels. Thus, a depth first search traversal will result in suffixes in lexicographical order.

## 3. RELATED WORK

This section presents the most important suffix tree construction algorithms classified into three main categories: *in-memory*, *semi-disk-based*, and *out-of-core*. Table 2 summarizes the comparison. It is worth to note that even though suffix trees are useful in a wide range of applications, there also exist specialized index structures for particular applications like genome data [11] and time series analysis [16].

**In-memory** approaches perform very well as long as the input string and the resulting suffix tree fit in main memory. This category includes algorithms, such as McCreight's



|  | **In-memory** | **Semi-disk-based** | | | | **Out-of-core** | |
| --- | --- | --- | --- | --- | --- | --- | --- |
| **Criteria** | **Ukkonen** | **Hunt** | **TDD** | **ST-Merge** | **TRELLIS** | **WaveFront** | **B$^2$ST** |
| Complexity | $O(n)$ | $O(n^2)$ | $O(n^2)$ | $O(n^2)$ | $O(n^2)$ | $O(n^2)$ | $O(n^2)$ |
| Memory locality | Poor | Good | Good | Good | Good | Good | Good |
| String access | Random | Random | Random | Random | Random | Sequential | Sequential |
| Parallel | No | No | No | No | No | Yes | No |

Table 2: Comparison of the most important algorithms for suffix tree construction

[12] and Ukkonen's [19]. For a string $S$ of size $n$, the time complexity of the latter is $O(n)$, which is optimal. However, this category suffers from poor locality of reference [18]. Once the suffix tree cannot fit in the main memory, the algorithms of this category require on average $O(n)$ expensive random disk I/Os. Recall that the suffix tree is an order of magnitude larger than the input string. Therefore, in practice in-memory methods are prohibitively expensive even for moderately long strings.

**Semi-disk-based** methods solve the locality of reference problem by decomposing the suffix tree into smaller sub-trees stored on the disk. This category includes Hunt's algorithm [10], TDD [17], ST-Merge [18] and TRELLIS [13]. The latter partitions the input string into several substrings, and constructs the corresponding sub-tree independently for each substring. The resulting sub-trees are stored on the disk. In a second phase, the sub-trees are merged into the final suffix tree. The time complexity is $O(n^2)$, but as long as the string $S$ fits into memory, the algorithms in this category perform few random I/Os so in practice they are faster than Ukkonen's algorithm. However, if $S$ is larger than the memory, the merging phase generates a lot of random disk I/Os rendering these algorithms very inefficient [2, 7]. It is worth noting that the sub-tree construction phase can be parallelizable but the merging phase is expected to require a lot of communication among processors. We are not aware of any parallel version of semi-disk-based algorithms.

**Out-of-core** category contains two recent methods that support strings larger than the main memory with reasonable efficiency by avoiding random I/Os. The first method, B$^2$ST [2] is based on suffix arrays [14]. A suffix array is a vector that contains all suffixes of the input string $S$ sorted in lexicographical order. A longest common prefix array is a vector that stores the length of the common prefix between each two consecutive entries in the suffix array. B$^2$ST divides the input string $S$ into several partitions and builds the corresponding suffix array and longest common prefix array for each partition. Then, it merges the suffix arrays of all partitions and generates suffix sub-trees. Note that the tree is constructed in batch at the final phase. This is an advantage of the algorithm because by avoiding the tree traversal for the insertion of each new node, it is more cache friendly. The time complexity is $O(cn)$, where $c = (2n/M)$ and $M$ is the size of the main memory. If $M$ is comparable to $n$ then $c$ is considered constant and the algorithm performs very well. However, as we mention in Section 1, in practice $n$ is expected to be much larger than $M$; in such a case the complexity becomes $O(n^2)$. A drawback of B$^2$ST is the large size of temporary results. The human genome for example is roughly 2.6G symbols whereas the temporary results are around 343GB. Furthermore, a parallel version of the algorithm would incur high communication cost among the processors during the merging phase; we are not aware of any parallel implementation.

WaveFront [7] is the second out-of-core algorithm. In contrast to B$^2$ST, which partitions the input string $S$, WaveFront works with the entire $S$ on independent partitions of the resulting tree $\mathcal{T}$. Tree partitioning is done using variable length S-prefixes (see example in Table 1), making sure that each sub-tree fits in main memory. Since $S$ may not fit in memory the algorithm may need to read $S$ multiple times. To minimize the I/O cost, WaveFront accesses $S$ strictly in sequential order. Each sub-tree is processed independently without a merging phase, so the algorithm is easily parallelizable. The parallel version has been implemented on an IBM BlueGene/L supercomputer; in absolute time it is the fastest existing method (it can index the human genome in 15 minutes [6]). Nevertheless, the algorithm cannot scale indefinitely, because more sub-trees increase the so-called *tiling overhead* [7]. Internally, WaveFront resembles the block nested loop join algorithm and requires two buffers. For optimum performance, these buffers occupy roughly 50% of the available memory, leaving the rest for the sub-tree. This is a drawback of the algorithm, because less memory leads to smaller and more trees that increase the tiling overhead. Moreover, even though the algorithm expands the sub-tree in layers, it needs to traverse the tree top-down for every new node, increasing the CPU cost.

Our approach, Elastic Range (ERA) is closer to Wave-Front, therefore there is no merging phase and it is easily parallelizable. However, ERA is significantly faster than WaveFront since it is based on properties that allow a level by level construction mechanism that performs clever memory management and minimizes the tiling overhead. Also, ERA amortizes the I/O cost by grouping together sub-trees where appropriate. Finally, ERA avoids multiple traverses of the sub-tree, achieving much lower CPU cost.

## 4. ELASTIC RANGE (ERA)

Elastic Range (ERA) is a novel approach that divides the problem vertically and horizontally (see Figure 3). Vertical partitioning splits the tree into sub-trees $\mathcal{T}_{p_1} \ldots \mathcal{T}_{p_n}$ that fit into the available memory using variable length S-prefixes similarly to [7, 10]. ERA goes a step further by grouping together sub-trees to share the I/O cost of accessing the input string $S$. Horizontal partitioning is applied independently in each sub-tree in a top-down fashion. The width of the horizontal partitions is adjusted dynamically (hence the name elastic range) based on how many paths in the sub-tree are still being processed. This allows ERA to use only a small part of the memory for buffers, rendering the algorithm cache-friendly and minimizing the tiling overhead. Each group represents an independent unit; groups can be processed serially or in parallel. The resulting sub-trees are assembled in the final suffix tree by a trie on the top. The trie is constructed with the S-prefixes used for vertical partitioning and is very small (e.g., the trie for the human genome is in the order of KB). The rest of this section describes the



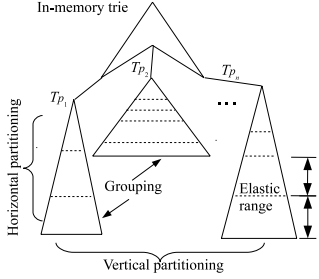

**Figure 3: Problem decomposition in ERa**

serial version of ERa. The parallel versions are discussed in Section 5.

## 4.1 Vertical Partitioning

Let $p$ be a S-prefix and $\mathcal{T}_p$ be the sub-tree that corresponds to $p$. Recall from Section 2 that $f_p$ is the number of suffixes with S-prefix $p$ (see example in Table 1). Each suffix corresponds to a leaf node in $\mathcal{T}_p$ and it is shown [10] that the number of internal nodes is equal to the number of leaves. The size in bytes of $\mathcal{T}_p$ is $2f_p \cdot sizeof(tree\_node)$. Let $MTS$ be the size of the memory reserved for the sub-tree. $\mathcal{T}_p$ can fit in the memory only if $f_p \leq \mathcal{F}_M$, where

$$\mathcal{F}_M = \frac{MTS}{2 \cdot sizeof(tree\_node)} \qquad (1)$$

To partition $\mathcal{T}$ into sub-trees that fit in $MTS$, we employ the idea of variable length S-prefixes [7, 10]. The algorithm starts by creating a working set containing one S-prefix for every symbol in the alphabet $\Sigma$. Then the entire input string $S$ is scanned to calculate the frequencies of each S-prefix in the working set. At the end of this step, each S-prefix whose frequency is at most $\mathcal{F}_M$ is removed from the working set. The remaining S-prefixes are extended by one symbol and the process is repeated until the working set is empty. In the example of Table 1, assume $\mathcal{F}_M = 5$. Since $f_{TG} = 7$, we extend $TG$ by one symbol and get $f_{TGA} = 1$, $f_{TGC} = 2$ and $f_{TGG} = 4$ that are all at most 5 and are removed from the working set; note that $f_{TGT} = 0$ since there is no $TGT$ substring in $S$. The worst case complexity is $\mathcal{O}(n^2)$ time, where $n$ is the size of $S$. In practice, for typical values of $MTS$ the algorithm runs in $\mathcal{O}(n)$. The human genome, for instance, requires 5 to 6 iterations when $MTS$ is 1 to 2GB.

However, the algorithm from [7] has a serious drawback: it generates unbalanced sub-trees that waste a lot of memory. In the previous example, the available memory can support frequencies up to $\mathcal{F}_M = 5$ but the frequencies of the resulting sub-trees are much smaller. Each small sub-tree is processed independently and accesses $S$ multiple times; therefore there are a lot of redundant I/Os. Also, a parallel implementation would waste resources because the CPUs that process the smaller sub-trees will be idle for long time. To avoid these problems, we propose a grouping phase after the initial partitioning.

We use a simple heuristic for grouping: The set of S-prefixes from the previous phase are put in a linked list sorted in descending frequency order. The head of the list (i.e., the S-prefix with the highest frequency) is added in a new group. Then, the list is traversed and S-prefixes are added to the group as long as the sum of the frequencies in the group is at most $\mathcal{F}_M$. The process is repeated until all S-

**Algorithm**: VERTICALPARTITIONING
**Input**: String $S$, alphabet $\Sigma$, $\mathcal{F}_M$ (see Equation 1)
**Output**: Set of $VirtualTrees$

1  $VirtualTrees := \varnothing$
2  $P := \varnothing$                    // linked list of S-prefixes
3  $P' := \{$ **for** every symbol $s \in \Sigma$ **do** generate a S-prefix $p_i = s$ $\}$
4  **repeat**
5      scan input string $S$
6      count in $S$ the appearances $f_{p_i}$ of every S-prefix $p_i \in P'$
7      **for** every $p_i \in P'$ **do**
8          **if** $0 < f_{p_i} \leq \mathcal{F}_M$ **then** add $p_i$ to $P$
9          **else for** every symbol $s \in \Sigma$ **do** add $p_i s$ to $P'$
10     remove $p_i$ from $P'$
11 **until** $P' = \varnothing$;
12 sort $P$ in descending $f_{p_i}$ order
13 **repeat**
14     $G := \varnothing$      // group of S-prefixes in a virtual tree
15     add $P.head$ to $G$ and remove the item from $P$
16     $curr :=$ next item in $P$
17     **while** $NOT$ end of $P$ **do**
18         **if** $f_{curr} + SUM_{g_i \in G}(f_{g_i}) \leq \mathcal{F}_M$ **then**
19             add $curr$ to $G$ and remove the item from $P$
20         $curr :=$ next item in $P$
21     add $G$ to $VirtualTrees$
22 **until** $P = \varnothing$;
23 **return** $VirtualTrees$

prefixes are processed (see Algorithm VERTICALPARTITIONING). In the previous example, this heuristic groups $TGG$ and $TGA$ together, whereas $TGC$ is in a different group. $TGG$ and $TGA$ share a common S-prefix $TG$ but this is a coincidence. The algorithm works with all S-prefixes generated from $S$ and may group together two or more completely unrelated S-prefixes.

A group of S-prefixes defines a *virtual* sub-tree that is processed as a single unit. When the input string $S$ is read from the disk, it is used by the entire group, therefore the I/O cost is amortized. Also, in a parallel environment, the utilization of resources is much better. Obviously, when $MTS$ is large, more sub-trees can be grouped together and the gain is larger.

## 4.2 Horizontal Partitioning

During this step ERa constructs the suffix sub-tree $\mathcal{T}_p$ for a S-prefix $p$ where $\mathcal{T}_p$ fits in the available main memory budget (Section 4.1). We base our method on properties of the suffix-tree (Proposition 1) that have not be exploited by previous approaches. Initially, in Section 4.2.1, we devise Algorithm COMPUTESUFFIXSUBTREE that exploits these properties to optimize access to the input string $S$. Then, Section 4.2.2 further extends the idea to also optimize main memory access (Algorithm SUBTREEPREPARE).

### 4.2.1 Optimizing String Access

To illustrate the key idea of our method, we will need the following notation. Let $e$ be an edge of $\mathcal{T}_p$. We denote by *(i)* $label(e)$ the label of $e$, *(ii)* $parent(e)$ the unique parent of $e$, and *(iii)* $pathlabel(e)$ the concatenation of edge labels on the path from the root to $e$. We consider nodes $u_1$, $u_2$ and $u_3$ of the suffix-tree illustrated in Figure 2 and make the following observations:

1. If an edge $e$ connects to a leaf then $pathlabel(e)$ appears only once in $S$. For instance, edge $e = (u_1, 14)$ that connects to leaf 14 has $pathlabel(e) = TGA \cdots \$$ that appears only once in $S$.

2. If an edge $e$ has a label of more than one symbols, say

**Algorithm**: COMPUTESUFFIXSUBTREE

**Input**: String $S$, S-prefix $p$
**Output**: The suffix sub-tree $\mathcal{T}_p(Nodes, Edges)$

1. $root := \text{new } Node(root)$
2. $u' := \text{new } Node$
3. $e' := \text{new } Edge(root, u')$
4. Label $e'$ with S-prefix $p$
5. BRANCHEDGE$(S, \mathcal{T}_p(Nodes, Edges), e')$
6. **return** $\mathcal{T}_p(Nodes, Edges)$

**Algorithm**: BRANCHEDGE

**Input**: String $S$, suffix sub-tree $\mathcal{T}_p(Nodes, Edges)$, edge $e(u_1, u_2)$

1. $Y$ is a set containing the symbols that follow $pathlabel(e)$ in $S$
2. **if** $pathlabel(e)$ appears once in $S$ **then**     // Leaf node
3.      Label $e$ with $label(e) \cdots \$$
4. **else if** $|Y| = 1$ **then** // Same symbol $s_1$ after $pathlabel(e)$ in $S$
5.      Extend the label of $e$ to include symbol $s_1$
6.      BRANCHEDGE$(S, \mathcal{T}_p(Nodes, Edges), e)$
7. **else for** each $s_i$ **do**
8.      $u' := \text{new } Node$
9.      $e' := \text{new } Edge(u_2, u')$
10.     Label $e'$ with $s_i$
11.     BRANCHEDGE$(S, \mathcal{T}_p(Nodes, Edges), e')$

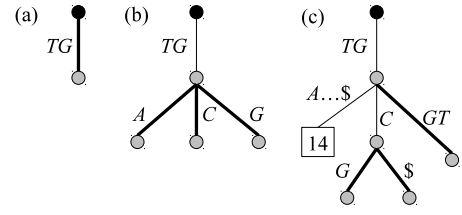

**Figure 4**: Constructing the suffix sub-tree of $TG$ (Example 1). Thin edges are finalized while thick edges need further processing.

$s_1 s_2 s_3 \cdots$, then $pathlabel(parent(e)) \cdot s_1$ is always followed by $s_2$ in $S$, $pathlabel(parent(e)) \cdot s_1 s_2$ is always followed by $s_3$ in $S$ and so on. For instance, for edge $e = (u_1, u_3)$ having label $GTG$, $TGG$ is always followed by $T$ in $S$ where $TG = pathlabel(parent(e))$.

3. If an edge $e$ is branched into another edge $e'$ then $pathlabel(e) \cdot s$, where $s$ is the first symbol of $label(e')$, appears (at least once) in $S$. For instance, edge $e = (u_1, u_2)$ is branched into edge $e' = (u_2, 9)$ and $TGCG$ appears in $S$, where (i) $TGC = pathlabel(e)$ and (ii) $G$ is the first symbol of $label(e')$.

Interestingly, the above observations are general properties of the suffix-tree that are captured formally by the following proposition.

PROPOSITION 1. *Let $S$ be a string and $e$ an edge of its suffix-tree.*

1. *Edge $e$ is connected to a leaf node iff $pathlabel(e)$ appears only once in $S$.*

2. *If $label(e) = s_1 \cdots s_k$, then substring $pathlabel(parent(e)) \cdot s_1 \cdots s_{i-1}$ is always followed by $s_i$ in $S$ $(1 < i \leq k)$.*

3. *Edge $e$ is branched into edges $e^1, \ldots, e^j$ iff $pathlabel(e) \cdot s^i$ $(1 \leq i \leq j)$ appears at least once in $S$ where $s^1, \ldots, s^j$ are distinct symbols formed by the first symbols of $label(e^1), \ldots, label(e^j)$ respectively.*

Contrary to previous suffix-tree construction approaches, Proposition 1 provides us with a method to build the suffix sub-tree $\mathcal{T}_p$ of a S-prefix $p$ level by level in a breadth-first fashion. This is achieved using Algorithms COMPUTESUFFIXSUBTREE and BRANCHEDGE. In more detail, Algorithm COMPUTESUFFIXSUBTREE creates an edge $e'$ labeled with $p$. Then, BRANCHEDGE is executed; it computes set $Y$ that stores all symbols appearing after $p$ in $S$. Following, the algorithm considers the 3 cases identified by Proposition 1. An example is shown below:

EXAMPLE 1. *Let us construct, using Algorithm COMPUTESUFFIXSUBTREE, the suffix sub-tree $\mathcal{T}_{TG}$ of S-prefix $TG$ for the string $S$ presented in Figure 2. In each step of the algorithm, we illustrate in Figure 4 the constructed suffix-tree.*

*Initially, Algorithm COMPUTESUFFIXSUBTREE creates an edge labeled with $TG$ and executes Algorithm BRANCHEDGE (Figure 4(a)). Since $Y = \{A, C, G\}$, Algorithm BRANCHEDGE branches the current edge into 3 new edges labeled with $A$, $C$ and $G$ (Figure 4(b)) and is recursively executed for these new edges. While processing these edges, the algorithm determines that (a) the edge labeled with $A$ connects to a leaf with label 14 (i.e., the offset of S-prefix $TGA \ldots \$$ in $S$), (b) the edge labeled with $C$ should be branched into two new edges labeled with $G$ and $\$$ respectively and (c) the edge labeled with $G$ should be extended to include symbol $T$ (Figure 4(c)). Algorithm BRANCHEDGE proceeds in a similar manner until sub-tree $\mathcal{T}_p$ is created.*

The heart of the suffix sub-tree construction mechanism lies in Algorithm BRANCHEDGE. For the clarity of presentation, this algorithm is illustrated in its simplest form (i.e., recursive) and without any optimization. The most costly operation of Algorithm BRANCHEDGE is the construction of set $Y$ since it requires a complete scan of the input string $S$. Such a scan is required for every modified edge. For instance, in Figure 4(b) BRANCHEDGE scans $S$ three times, one for each thick edge (labeled with $A$, $C$ and $G$). Also, for each scan the algorithm reads and stores in $Y$ only one symbol after each occurrence of $pathlabel(e)$ in $S$.

The actual implementation of BRANCHEDGE is iterative (non recursive) and has three major optimizations:

1. The cost of scanning $S$ is amortized for all the edges of a level. For Figure 4(b), a single scan of $S$ is needed to process all thick edges $A$, $C$ and $G$.

2. For each scan of $S$, the algorithm reads a range of symbols. This means that $Y$ is now a set of strings (instead of a set of symbols). The exact size of the range depends on the available main memory (see Section 4.4). In total, by reading $l$ symbols, we reduce the scans of $S$ by a factor $l$.

3. The algorithm constructs the sub-tree $\mathcal{T}_p$ for a S-prefix $p$. If more sub-trees are grouped in the same virtual tree (see Section 4.1), then each scan of $S$ updates the edges of all sub-trees in the virtual tree.

From this point onwards, we consider that the COMPUTESUFFIXSUBTREE algorithm uses the above described optimized version of BRANCHEDGE.

### 4.2.2 Optimizing Memory Access

The experimental evaluation and profiling of the Algorithm COMPUTESUFFIXSUBTREE showed that a significant



amount of time is spent on updating the constructed (in main memory) suffix sub-tree $\mathcal{T}_p$. This is mainly due to the fact that the construction process requires memory accesses that may not be sequential nor local. To address this issue, we propose a novel two step approach formed by a *preparation* and a *construction* step.

**The preparation step** is executed by Algorithm SUB-TREEPREPARE. It extends the optimized version of BRANCH-EDGE and employs Proposition 1 to construct *not* the sub-tree but a novel intermediate data structure.

**The construction step** is performed by Algorithm BUILD-SUBTREE that utilizes the data structure produced by the preparation step to construct the suffix sub-tree in batch. By decoupling the sub-tree construction from the preparation, we localize memory accesses and avoid costly traversals of the partial sub-tree for each new node.

The crux of the proposed method is the intermediate data structure. It consists of array **L** that stores the leaves and array **B** that stores branching information. More precisely, array **L** stores the positions of the input S-prefix $p$ in $S$, i.e., the leaves of the sub-tree. The order of the leaves in **L** is such that the corresponding suffixes are lexicographically sorted, i.e., $S_{L[i]} \leq S_{L[i+1]}$. Array **B** is more involved and consists of triplets of the form $(c_1, c_2, offset)$ where $c_1$ and $c_2$ are symbols and $offset$ is an integer. Intuitively, **B**[i] describes the relation between the branch $Br_{i-1}$ that leads to **L**[i − 1] and the branch $Br_i$ that leads to **L**[i] ($1 \leq i$). Specifically, $offset$ is the number of symbols in the common path of $Br_{i-1}$ and $Br_i$ (this corresponds to the size of the common longest S-prefix of $S_{\mathbf{L}[i-1]}$ and $S_{\mathbf{L}[i]}$). Symbol $c_1$ (respectively $c_2$) is the first symbol of the branch to **L**[i − 1] (respectively **L**[i]) after their separation. For instance, using **L** and **B** we can represent $\mathcal{T}_{TG}$ (Figure 2) as follows:

| | 0 | 1 | 2 | 3 | 4 | 5 | 6 |
|---|---|---|---|---|---|---|---|
| **L** | 14 | 9 | 20 | 6 | 17 | 3 | 0 |
| **B** | | (A,C,2) | (G,$,3) | (C,G,2) | (G,$,6) | (C,G,5) | (C,G,8) |

For example, *(a)* **L**[0] = 14 since the lexicographically smallest suffix is $S_{14}$ (Table 1), and *(b)* **B**[5] = (C, G, 5) since the branch leading to **L**[4] = 17 separates after 5 symbols (i.e., $TGGTG$) from the branch leading to **L**[5] = 3 and $C$, $G$ are the first symbols after the separation (Figure 2).

To compute arrays **L** and **B** that correspond to the sub-tree $\mathcal{T}_p$ of a S-prefix $p$, we employ Algorithm SUBTREE-PREPARE (recall that **L** stores the leaves and **B** the branching information of $\mathcal{T}_p$). The algorithm uses 4 auxiliary data structures of size |**L**|, namely **I**, **A**, **R** and **P**. During the process the order of the elements in **A**, **R**, **P** and **L** may change. Intuitively:

**Array R** is the main memory buffer of the input string $S$. Specifically, **R**[i] stores symbols required to construct the branch leading to leaf **L**[i].

**Array I** is an index that holds information that restores the original order of leaves in the string $S$. More precisely, the position of the $i$th leaf in $S$ may be accessed using $\mathbf{L}\big[\mathbf{I}[i-1]\big]$. If **I**[i] = *done* then the corresponding branch is completed. In other words, leaves $\mathbf{L}\big[\mathbf{I}[0]\big], \ldots, \mathbf{L}\big[\mathbf{I}[|\mathbf{L}|-1]\big]$ appear in that order in $S$. Thus, to fill **R**, $S$ is sequentially read until the symbols pertaining to leaves $\mathbf{L}\big[\mathbf{I}[0]\big], \ldots, \mathbf{L}\big[\mathbf{I}[|\mathbf{L}|-1]\big]$ are found and stored in $\mathbf{R}\big[\mathbf{I}[0]\big], \ldots, \mathbf{R}\big[\mathbf{I}[|\mathbf{L}|-1]\big]$, respectively (Lines 10-12 of SUBTREEPREPARE). Overall, array **I** is of paramount importance since it allows us to fill the buffers

**Algorithm**: SUBTREEPREPARE
**Input**: Input string $S$, S-prefix $p$
**Output**: Arrays **L** and **B** corresponding suffix sub-tree $\mathcal{T}_p$

1 **L** contains the locations of S-prefix $p$ in string $S$
2 **B** := ()
3 **I** := $(0, 1, \ldots, |\mathbf{L}| - 1)$
4 **A** := $(0, 0, \ldots, 0)$
5 **R** := ()
6 **P** := $(0, 1, \ldots, |\mathbf{L}| - 1)$
7 $start := |p|$              // Start after S-prefix $p$
8 **while** *there exist an undefined* **B**[i], $1 \leq i \leq |\mathbf{L}| - 1$ **do**
9     $range :=$ GETRANGEOFSYMBOLS       // Elastic range
10     **for** $i := 0$ **to** $|\mathbf{L}| - 1$ **do**
11         **if** **I**[i] $\neq$ *done* **then**
12             $\mathbf{R}\big[\mathbf{I}[i]\big] :=$ READRANGE$(S, \mathbf{L}\big[\mathbf{I}[i]\big] + start, range)$
            // READRANGE$(S, a, b)$ reads $b$ symbols of $S$
            starting at position $a$
13     **for** *every active area AA* **do**
14         Reorder the elements of **R**, **P** and **L** in $AA$ so that **R** is lexicographically sorted. In the process maintain the index **I**
15         If two or more elements $\{a_1, \ldots, a_t\} \in AA$, $2 \leq t$, exist such that $\mathbf{R}[a_1] = \cdots = \mathbf{R}[a_t]$ introduce for them a new active area
16     **for** *all i such that* **B**[i] *is not defined*, $1 \leq i \leq |\mathbf{L}| - 1$ **do**
17         $cs$ is the common S-prefix of **R**[i − 1] and **R**[i]
18         **if** $|cs| < range$ **then**
19             $\mathbf{B}[i] := (\mathbf{R}[i-1][|cs|], \mathbf{R}[i][|cs|], start + |cs|)$
20             **if** **B**[i − 1] *is defined or* $i = 1$ **then**
21                 Mark $\mathbf{I}\big[\mathbf{P}[i-1]\big]$ and **A**[i − 1] as *done*
22             **if** **B**[i + 1] *is defined or* $i = |\mathbf{L}| - 1$ **then**
23                 Mark $\mathbf{I}\big[\mathbf{P}[i]\big]$ and **A**[i] as *done* // Last element of an active area
24     $start := start + range$
25 **return** (**L**, **B**)

of **R** in a single sequential scan of $S$ (and thus retain the properties of Algorithm BRANCHEDGE).

**Array A** identifies the active areas of the process. Elements $i$ and $i + 1$ belong to the same active area if **A**[i] = **A**[i+1]. If **A**[i] = *done* then element $i$ is completed.

**Array P** stores the order of appearance in the string $S$ of the leaves in **L**. If **P**[i] = $x$ then leaf **L**[i] corresponds to the $x + 1$ appearance of S-prefix $p$ in $S$. **P** is used in Lines 21 and 23.

We will illustrate Algorithm SUBTREEPREPARE using the following example.

EXAMPLE 2. *We will construct arrays **L** and **B** of the suffix sub-tree $\mathcal{T}_{TG}$ of S-prefix TG for the string S presented in Figure 2. The algorithm starts by initializing all necessary structures* (**I**, **A**, **R**, **L** *and* **B**). *Then, the algorithm decides to read ranges of 4 symbols[2] from the input string S (range =* 4*) to fill the buffers of **R** (Lines 9-12). The values of the variables up to this point are as follows:*

**Trace 1**

| | 0 | 1 | 2 | 3 | 4 | 5 | 6 |
|---|---|---|---|---|---|---|---|
| **I** | 0 | 1 | 2 | 3 | 4 | 5 | 6 |
| **A** | 0 | 0 | 0 | 0 | 0 | 0 | 0 |
| **R** | GTGG | GTGG | GTGC | CGGT | ATGG | GTGC | C$ |
| **P** | 0 | 1 | 2 | 3 | 4 | 5 | 6 |
| **L** | 0 | 3 | 6 | 9 | 14 | 17 | 20 |

*For instance, if $i = 3$ the algorithm considers position* **I**[3] = 3 *and reads from S symbols CGGT that correspond*

---
[2] We will discuss how the range is determined in Section 4.4



*to the range = 4 symbols after position* $\mathbf{L}[\,\mathbf{I}[i]\,] + start = 9 + 2 = 11$.

*Following (Lines 13-15), the algorithm considers all elements (since they belong to the same active area marked with 0) and reorders* $\mathbf{R}$, $\mathbf{P}$ *and* $\mathbf{L}$ *so that* $\mathbf{R}$ *is lexicographically sorted, while maintaining array* $\mathbf{I}$. *The algorithm continues executing Lines 16-23 which compute array* $\mathbf{B}$. *The results of this iteration are illustrated below:*

**Trace 2**

|   | 0 | 1 | 2 | 3 | 4 | 5 | 6 |
|---|---|---|---|---|---|---|---|
| **I** | 5 | 6 | 3 | done | done | 4 | done |
| **A** | done | done | done | 1 | 1 | 2 | 2 |
| **R** | ATGG | CGGT | C$ | GTGC | GTGC | GTGG | GTGG |
| **P** | 4 | 3 | 6 | 2 | 5 | 0 | 1 |
| **L** | 14 | 9 | 20 | 6 | 17 | 0 | 3 |
| **B** |  | (A,C,2) | (G,$,3) | (C,G,2) |  | (C,G,5) |  |

*Note that Lines 13-15 place at the fourth position ($i = 3$) leaf* $\mathbf{L}[3] = 6$ *that corresponds to the lexicographically fourth suffix of TG (i.e., $S_6$). The fact that the current position (3) was moved from position 2 of the initial order is marked by* $\mathbf{I}[2] = 3$. *Also, Line 15 identifies two more active areas denoted by 1 and 2 in* $\mathbf{A}$.

*Also note that, for $i = 1$, Lines 16-23 focus on* $\mathbf{R}[0] = ATGG$ *and* $\mathbf{R}[1] = CGGT$, *which do not have a common S-prefix (i.e, $|cs| = 0$). Thus, the algorithm sets (a) $\mathbf{B}[1] = (\mathbf{R}[0][0], \mathbf{R}[i][0], 2 + 0) = (A, C, 2)$ and (b) $\mathbf{I}[\mathbf{P}[0]] = \mathbf{I}[4] = done$ and $\mathbf{A}[0] = done$. The equation in (a) illustrates that sub-tree branches after $start + |cs| = 2$ symbols (i.e., TG) and follows from Proposition 1, Case 3. The equations in (b) show that suffix $S_{\mathbf{L}[0]} = TGATGG \cdots \$$, that appears only once in S, does not need any further processing (follows from Proposition 1, Case 1).*

*In each iteration, the construction of* $\mathbf{B}$ *involves sequential access of array* $\mathbf{R}$ *and can be performed very efficiently by a single memory scan. The next (and final) iteration of the while loop (Lines 8-24) considers only $i \in \{0, 1, 2, 5\}$ for which $\mathbf{I}[i] \neq done$ and fills only the appropriate elements of* $\mathbf{R}$ *(5, 6, 3, 4 respectively). After the execution of Lines 13-23, the structures are update as follows:*

**Trace 3**

|   | 0 | 1 | 2 | 3 | 4 | 5 | 6 |
|---|---|---|---|---|---|---|---|
| **I** | done | done | done | done | done | done | done |
| **A** | done | done | done | done | done | done | done |
| **R** |  |  |  | GGTG | $ | TGCG | TGGT |
| **P** | 4 | 3 | 6 | 2 | 5 | 0 | 1 |
| **L** | 14 | 9 | 20 | 6 | 17 | 3 | 0 |
| **B** |  | (A,C,2) | (G,$,3) | (C,G,2) | (G,$,6) | (C,G,5) | (C,G,8) |

*Note the reorder of the elements of* $\mathbf{R}$, $\mathbf{P}$ *and* $\mathbf{L}$ *for $i \in \{5, 6\}$ that correspond to the lexicographical sorting of active area tagged with 2.*

Summarizing, Algorithm SUBTREEPREPARE retains the sequential access of the input string $S$ (using array $\mathbf{I}$) but also constructs $\mathbf{I}$ and $\mathbf{B}$ using sequential main memory access. Algorithm BUILDSUBTREE takes these structures and builds the corresponding suffix sub-tree using also sequential memory access.

EXAMPLE 3. *We continue Example 2. Algorithm BUILDSUBTREE creates an edge that links the root with the lexicographically first leaf* $\mathbf{L}[0] = 14$. *This edge is labeled with*

**Algorithm**: BUILDSUBTREE
**Input**: Arrays $\mathbf{L}$ and $\mathbf{B}$
**Output**: The corresponding suffix sub-tree $\mathcal{T}_p$

1  $root :=$ new $Node(root)$
2  $u' :=$ new $Node$
3  $e' :=$ new $Edge(root, u')$
4  Label $e'$ with $S_{\mathbf{L}[0]}$    // The suffix that corresponds $\mathbf{L}[0]$
5  Label $u'$ with $\mathbf{L}[0]$       // First (lexicographically) leaf
6  Push $e'$ to $Stack$
7  $depth := |label(e')|$
8  **for** $i := 1$ **to** $|\mathbf{B}| - 1$ **do**
9      $(c_1, c_2, \textit{offset}) := \mathbf{B}[i]$
10     **repeat**
11        Pop an edge $se(v_1, v_2)$ from the $Stack$
            $depth := depth - |label(se)|$
12     **until** $depth \leq \textit{offset}$;
13     **if** $depth = \textit{offset}$ **then**
14        $u := v_1$
15     **else**
16        Break edge $se(v_1, v_2)$ into edges $se_1(v_1, v_t)$ and $se_1(v_t, v_2)$
17        Label $se_1$ with the first $\textit{offset}$ symbols of $label(se)$
18        Label $se_2$ with the remaining symbols
19        $u := v_t$
20        Push $se_1$ to $Stack$
21        $depth := depth + |label(se_1)|$
22     $u' :=$ new $Node$
23     $ne :=$ new $Edge(u, u')$
24     Label $ne$ with $S_{\mathbf{L}[i]}$    // The suffix that corresponds $\mathbf{L}[i]$
25     Label $u'$ with $\mathbf{L}[i]$        // Next (lexicographically) leaf
26     Push $ne$ to $Stack$
27     $depth := depth + |label(ne)|$
28 **return** $\mathcal{T}_p$

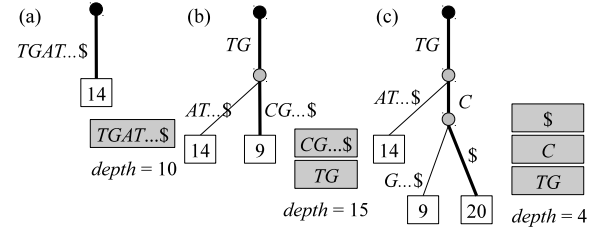

**Figure 5:** Trace of the BuildSubTree algorithm for the first four entries of Trace 3. The stack is depicted in gray next to each sub-tree

$TGAT \cdots \$$, *which is the suffix that corresponds to* $\mathbf{L}[0]$. *Also this edge is pushed to the stack and variable depth is initialized to 10 (i.e., the size of the label). All the above are illustrated in Figure 5(a). Then, the algorithm goes through the following iterations:*

**The $1^{st}$ iteration** *considers* $\mathbf{B}[1] = (c_1, c_2, \textit{offset}) = (A, C, 2)$. *Edge $TGAT \cdots \$$ is popped. Since offset = 2, this edge breaks into two edges labeled with TG and $AT \cdots \$$ (Lines 15-21). Moreover, a new edge is created that (a) links edge TG and the second leaf $\mathbf{L}[1] = 9$ and (b) is labeled with $CG \cdots \$$. Also, edges TG and $CG \cdots \$$ are pushed to Stack and depth = $|label(TG)| + |label(CG \cdots \$)| = 2 + 13 = 15$. This iteration is depicted in Figure 5(b).*

**The $2^{nd}$ iteration** *considers* $\mathbf{B}[2] = (G, \$, 3)$ *and proceeds in a similar manner. It is depicted in Figure 5(c).*

*The remaining iterations are similar and are omitted.*

### 4.3 Complexity Analysis

Let $S$ be the input string, $n = |S|$ be its length, $LP$ be the longest path label in the suffix tree of $S$ and $\mathbf{L}$ be the largest



list that stores the offset of each occurrence of a S-prefix $p$.
In the worst case, $|LP| = \mathcal{O}(n)$ and $|\mathbf{L}| = \mathcal{O}(n)$. To see
this, consider $S = aaaa\$$ for which $n = 4$, $|LP| = 3$ (since
S-prefix $aaa$ appears at positions 0 and 1) and $|\mathbf{L}| = 4$ (since
S-prefix $a$ appears 4 times).

Algorithm SUBTREEPREPARE in each iteration of the while
loop (Lines 8-24) retrieves $range$ symbols for each entry of
$\mathbf{L}$ and sorts them lexicographically. Thus, each iteration
takes $|\mathbf{L}|\log|\mathbf{L}|$ time. Moreover, each iteration is performed
$\frac{|LP|}{range}$ times. Therefore, the overall worst case complexity of
SUBTREEPREPARE is $|\mathbf{L}| \cdot \log|\mathbf{L}| \cdot \frac{|LP|}{range}$ which is $\mathcal{O}(n^2 \log n)$
time. Algorithm BUILDSUBTREE generates one leaf node
for each entry of $\mathbf{L}$. To this end, it accesses the stack up to
$|LP|$ times. Therefore, its worst case complexity is $|\mathbf{L}| \cdot |LP|$
which is $\mathcal{O}(n^2)$.

However, in practice and in all application scenarios $\mathbf{L} \ll n$ and $|LP| \ll n$ hold. In fact it is reasonable to expect that
$\mathbf{L}$ and $|LP|$ are orders of magnitude smaller than $n$. Thus,
the overall expected complexity bound of ERA is much better than the worst case bound. This is also verified by the experimental evaluation, which demonstrated that ERA scales almost linearly to $n$.

## 4.4 Memory Allocation and Disk Access

Efficient allocation of the available memory is critical because, if more memory is available for the sub-tree, vertical partitioning will generate fewer virtual trees, hence the I/O cost will be lower. Let $MTS$ be the maximum tree size and $f_p = |\mathbf{L}|$ be the frequency of the sub-tree $\mathcal{T}_p$ of S-prefix $p$. Recall from Section 4.1 that $|\mathbf{L}| \leq \mathcal{F}_M$. ERA divides the available memory into three parts (see Figure 6):

**Retrieved data area.** It contains the input buffer **BS** and the array **R** of next symbols. It also contains a small area (less than 1MB) for the trie that connects sub-trees.

**Processing area.** It contains data structures that are used during construction. These include arrays **I**, **L**, **P**, **A** and **B**. The size of all of these arrays is of factor $|\mathbf{L}|$. **L** together with **B** consume almost 40% of the available memory.

**Suffix tree area.** Its size (i.e., $MTS$) is roughly 60% of the total available memory.

The size of **BS** is relatively small and should be a multiple of the block size of the underlying I/O subsystem; in our environment 1MB was adequate. The size of **R** affects the range of symbols to be fetched in each scan (Line 5, Algorithm SUBTREEPREPARE). A large **R** minimizes the number of string scans while a small **R** avoids unnecessary reads and frequent cache misses. These occur when algorithm reads $range$ symbols from $S$ but only few of them are needed to determine that it corresponds to a leaf and does not need further processing. The proper size of **R** mainly depends on the alphabet size, which determines the branching factor of the tree. Intuitively, to build suffix trees with a larger branching factor, we require more concurrent active areas and thus a larger size of **R**. In our experiments, we found that a good size for small alphabets (e.g, DNA data) is 32MB whereas for large alphabets (e.g., Protein data) it should be 256MB (Figure 8).

Observe that the processing and the suffix tree areas in Figure 6 overlap. SUBTREEPREPARE uses part of the suffix tree area to store arrays **I**, **A** and **P**. Recall that the sub-tree is constructed in batch by Algorithm BUILDSUBTREE which only needs arrays **L** and **B**; therefore, **I**, **A** and **P** can be safely overwritten.

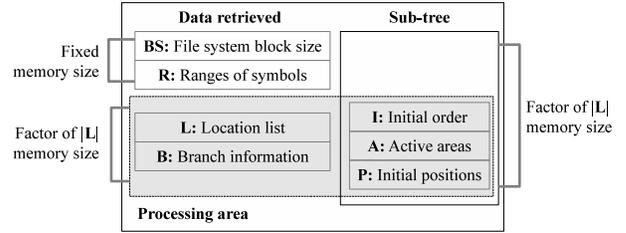

**Figure 6: Allocation of the available memory. The processing and sub-tree area overlap**

ERA implements dynamic memory management to reduce significantly the I/O cost. Recall that Algorithm SUBTREE-PREPARE scans the string $\frac{|LP|}{range}$ times. While the size of **R** is constant, the number of active areas in **L** are reduced after each iteration if new leaves are discovered; inactive areas do not need space in **R**. Let $|\mathbf{L}'| \leq |\mathbf{L}|$ be the number of **L** entries that belong to active areas at the current iteration. Line 9 in SUBTREEPREPARE calculates the range of next symbols to prefetch as $range = \frac{|\mathbf{R}|}{|\mathbf{L}'|}$. In practice, after a few scans of $S$ a lot of areas become inactive and $range$ becomes large, leading to a dramatic improvement in I/O cost. The gain becomes more significant as the size of the input increases. Our experiments revealed that, for very long strings, the performance of the algorithm is doubled.

ERA also optimizes the disk access pattern. Previous methods (e.g., WaveFront) at each scan read the entire string in sequential order. The intuition is that *(a)* sequential order avoids the seek time, therefore it is roughly an order of magnitude faster than random I/Os in modern disks; and *(b)* since the probability of finding at least one required symbol within each disk block is high, only a few blocks will be unnecessarily fetched if the entire $S$ is read. While *(b)* is true for the initial iterations of ERA, we observed that, as more leaves are discovered and areas become inactive, the probability of fetching a block that does not contain any required symbol increases significantly. For this reason, we implemented a simple heuristic: If a block (or a continuous range of blocks) is not expected to contain any necessary symbol, we skip these blocks by performing a random seek. Whether next block(s) contain at least one necessary symbol can be determined by the information in **I** and $range$. Note that, even though a random seek is performed, the seek time is expected to be very small because the next block is physically very close to the current disk head position. The experiments show a gain of up to 10%.

## 5. PARALLEL CONSTRUCTION

Indexing very long strings can use parallel computing resources and aggregated CPU power to achieve better performance. Most existing suffix tree construction algorithms (including the recent B$^2$ST) are not easily parallelizable because each thread processes a small portion of the string $S$ and there is a costly phase that merges thread results. To the best of our knowledge, the most successful parallel construction algorithm is PWaveFront [6].

Horizontal partitioning of ERA is easily parallelizable because each process is independent and there is no merging phase (Section 4.2.2). We developed two parallel versions of ERa: one for shared-memory and shared-disk systems (e.g., typical multicore desktops) and a second one for shared-nothing architectures (e.g., computer clusters or cloud computing infrastructure). We did not parallelize the vertical partitioning phase since its cost is low.

**Shared-memory and shared-disk.** This category contains multicore systems where cores share the main system's RAM and disk. A master thread at one of the cores generates groups of variable length prefixes and divides these groups equally among the available cores including itself. The main advantage of this architecture is that the input string is available to all cores. A significant drawback is the bottleneck at the memory bus and I/O subsystem when multiple cores attempt to access the string. Therefore, scalability is expected to be limited.

**Shared-nothing architecture.** In this architecture, each node has its own disk and memory; thus the aggregated I/O and memory bandwidth scale with the number of nodes. Again, a master node generates groups of variable length prefixes and divides them equally among the available nodes including itself. Since each node works independently, this architecture has the potential to scale-up very well. Note, however, that during initialization the input string should be transmitted to each node; this is the main drawback of this architecture. We expect that the problem can be minimized by using an appropriate parallel file system.

## 6. EXPERIMENTAL EVALUATION

This section presents the performance evaluation for the serial and parallel versions of ERa. We compare our work against the two existing *out-of-core* approaches, $B^2ST$ and WaveFront, and a *semi-disk-based* approach: Trellis. For $B^2ST$ [2] and Trellis [13], we downloaded the serial implementation from the authors' sites. There is no parallel version and the existing implementations support only strings with 4 symbols. WaveFront was not available, so we implemented our own serial version following [7] and a parallel version following PWaveFront [6].

We used large real datasets: *(a)* The Human Genome[3] with size roughly 2.6GBps[4] and alphabet of 4 symbols; *(b)* DNA[5], containing 4GBps from an alphabet of 4 symbols, which is the concatenation of horse, zebra-fish and human DNA sequences; *(c)* the Protein[6] dataset containing 4GBps from an alphabet of 20 symbols, and *(d)* the English text from Wikipedia[7] containing 5G characters from an alphabet of 26 symbols.

### 6.1 Serial Version

All serial methods were implemented in C, except Trellis that was implemented in C++, and compiled with gcc version 4.4.1 in Linux. The experiments were executed on a machine with two quad-core Intel CPUs at 2.67GHz and 24GB RAM. As the main focus is on *out-of-core* approaches, our experiments used a ratio of memory budget to input string size that is up to 1:5. We limited the available mem-ory using `ulimit -v`, and turned off the virtual memory (`swapoff -a`).

---
[3] http://webhome.cs.uvic.ca/~thomo/HG18.fasta.tar.gz
[4] GBps: Giga Base pairs - equivalent to $10^9$ symbols
[5] http://www.ensembl.org/info/data/ftp/index.html
[6] http://www.uniprot.org/uniprot/?query=&format=*
[7] http://en.wikipedia.org/wiki/Wikipedia:Database_download

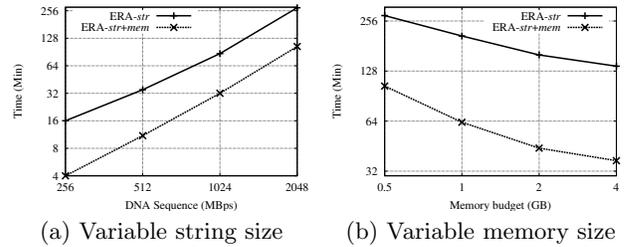

(a) Variable string size    (b) Variable memory size

**Figure 7:** Serial execution time of horizontal partitioning methods; DNA dataset. (a) 512MB RAM; (b) $|S|$=2GBps

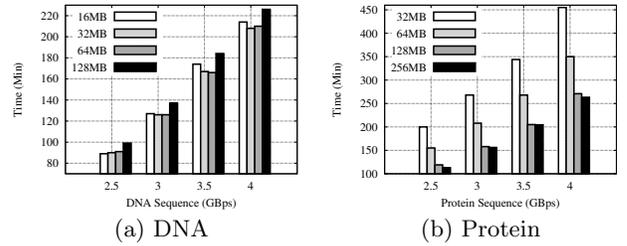

(a) DNA    (b) Protein

**Figure 8:** Tuning the size of R: 32MB for DNA ($|\Sigma|=4$); 256MB for Protein ($|\Sigma|=20$)

**Horizontal partitioning.** In this experiment, we compare the two horizontal partitioning methods proposed in Section 4.2 for ERa. The 1st approach (ERa-*str*) uses Algorithms ComputeSuffixSubTree and BranchEdge that tune string access (Section 4.2.1), while the 2nd approach (ERa-*str+mem*) uses Algorithms SubTreePrepare and BuildSubTree that tune string and memory access (Section 4.2.2). We varied the size of the input string from 256MBps to 2048MBps (DNA dataset) while setting the memory size to 512MB, as shown in Figure 7(a). Moreover, Figure 7(b) compares the construction time for a 2GBps DNA sequence with memory size varying from 0.5 to 4GB. These experiments demonstrate that ERa-*str+mem* (Section 4.2.2) offers significant improvements.

**ERa tuning.** Here we tune the size of **R** (i.e., read-ahead buffer for next symbols) that significantly affects the performance of ERa (see also Section 4.4). Larger **R** means less scans of the string while smaller **R** avoids unnecessary reads. Since the size of **R** depends on the size of the alphabet, we seek the proper size of **R** for a small alphabet (DNA dataset of 4 symbols) and a large alphabet (Protein dataset of 20 symbols). To this end, we fix the memory to 1GB and generate inputs with 2.5 to 4GBps from the prefixes of DNA and Protein datasets. Figure 8(a) shows that 32MB is a good value for the DNA dataset, whereas Figure 8(b) shows that 256MB is appropriate for the Protein dataset, which has larger alphabet. The results for English were similar to Protein, since both datasets need 5 bits to encode each symbol; therefore, we used $|\mathbf{R}|$=256MB for English, too.

Unless otherwise mentioned, the following experiments use the disk seek optimization, described in Section 4.4. This optimization improved the performance of the serial version

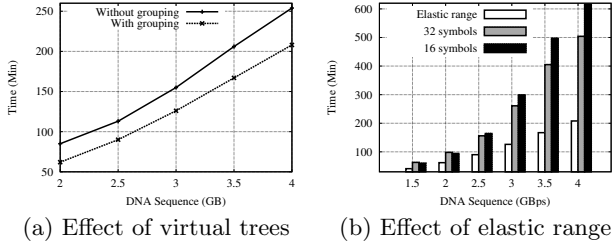

(a) Effect of virtual trees    (b) Effect of elastic range

**Figure 9: Effect of virtual trees and elastic range. DNA dataset; 1GB RAM; serial execution**

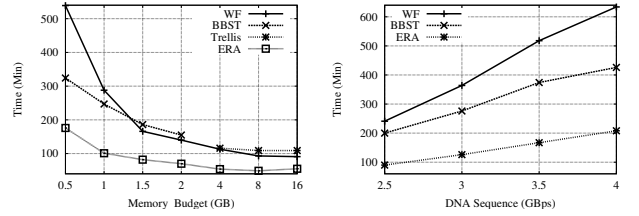

(a) Variable memory; Human Genome    (b) Variable string size; DNA; 1GB RAM

**Figure 10: Serial execution time versus (a) available memory and (b) string size**

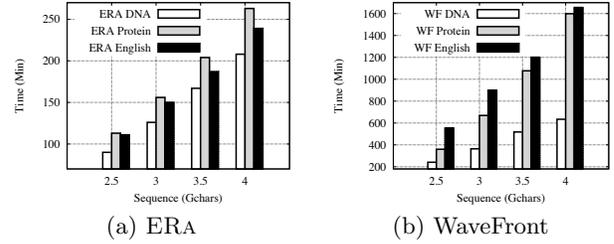

(a) ERa    (b) WaveFront

**Figure 11: Serial execution time of (a) ERa and (b) WaveFront versus variant alphabets; 1GB RAM**

by roughly 10% (see Section 6.2 for more details).

**Vertical partitioning.** Recall from Section 4.1 that vertical partitioning in ERa extends WaveFront by grouping the sub-trees into virtual trees to amortize the I/O cost. Figure 9(a) compares the effect of using the virtual trees versus no tree grouping, for the DNA dataset and 1GB RAM. Virtual trees achieve at least 23% better overall performance.

**Elastic range.** The next experiment shows the effect of the elastic range approach. Recall from Section 4.4 that as more areas become inactive, ERa uses the space that becomes available in **R** (whose size is constant) to prefetch more symbols for the active areas. Figure 9(b) compares the performance of elastic range against two alternatives that use static ranges of 16 and 32 prefetched symbols; elastic range is 46% to 240% faster and the gain increases for very long strings. Note that using a larger static range is not a good alternative to the elastic range. For example, 32 symbols is 22% faster than 16 symbols for string size equal to 4GBps, but it is 13% slower than 16 symbols for $|S| =$ 1.5GBps.

**Comparison against WaveFront, B$^2$ST and Trellis.** The following experiments compare ERa against WaveFront and B$^2$ST. We have allotted to all algorithms the same amount of memory. B$^2$ST allocates the memory to the input and output buffers and the intermediate data, such as suffix arrays. For WaveFront, the best setting according to [7] divides the memory equally between the processing space, the input buffers and the sub-tree. In contrast, ERa first allocates memory for **R** (according to Figure 8), 1MB for the input buffer and 3MB for the trie index. 60% of the remaining memory is allocated to the sub-tree and the rest is used for processing space (i.e., arrays **B** and **L**). **A**, **P** and **I** are located temporally in the area of the sub-tree, as discussed in Section 4.4. Because of the better allocation, ERa can construct larger sub-trees than WaveFront using the same amount of memory.

Figure 10(a) compares the construction time for the Human Genome dataset with memory size ranging from 0.5 to 16GB. ERa is consistently twice as fast compared to the best competitor, where string size is larger than the memory budget (out-of-core construction). It is worth noting that, while WaveFront is slightly faster than B$^2$ST for large memory size, it is dramatically slower when the memory is limited. Note that, the available implementation of B$^2$ST does not support large memory; this is why B$^2$ST plot stops at 2GB.

We also compared the performance of ERa against the performance of WaveFront and Trellis using large memory budget. Note that Trellis needs to accomodate the entire input string in memory. Since the human genome cannot fit in 2GB memory, the plots for Trellis start at 4GB. Recall that both ERa and WaveFront access the string sequentially from disk during construction. As shown in Figure 10(a), both ERa and WaveFront outperform Trellis. Although Trellis does not pay the I/O cost of accessing the string, it has to access in random fashion the large sub-trees (total size is roughly 26 times larger than the input string) from the disk during the merging phase. Our results agree with those from [2, 7].

Furthermore, we varied the size of the input string from 2.5 to 4GBPS (DNA dataset) while setting the memory size to 1GB. The total execution time is shown in Figure 10(b). ERa is at least twice as fast as its competitors. The performance gap from WaveFront is increasing for longer strings.

Finally, we evaluated the performance of ERa and WaveFront for different alphabet sizes. Figure 11 shows the results of both methods for DNA ($|\Sigma|=4$), Protein ($|\Sigma|=20$), and English ($|\Sigma|=26$) datasets. For ERa, since DNA has only 4 symbols, each symbol is encoded in 2 bits, in contrast to Protein and English that need 5 bits per symbol. Therefore, a larger part of the DNA string can fit in the memory. Also fewer symbols mean smaller branch factor, therefore less CPU cost. For these reasons using ERa, the processing of DNA is around 20% times faster than Protein and English. Moreover, the longest sequence that is repeated in the English dataset is shorter than the longest repeated sequence in the Protein dataset. The longest repeated sequence affects the depth of the tree. Therefore, ERa indexes the English dataset faster than the Protein.

WaveFront inserts the suffixes (leaf nodes) ordered as they appear in the string from left to right. Since the leaves of the suffix tree are sorted lexicographically, nodes that are logically connected, are not physically nearby in the suffix tree built by WaveFront. The more symbols, the larger the branch factor, which leads to more random memory accesses. Consequently, WaveFront spends a lot of time navigating the tree in order to insert a new leaf node. In contrast, since ERA sorts first the leaves lexicographically, it is not affected significantly by the branch factor.

## 6.2 Parallel Version

We developed parallel versions of ERA and WaveFront using MPI[8]. For WaveFront we followed [6]. There is no existing parallel version of B$^2$ST. Moreover, such an implementation would probably be inefficient because of the costly merging phase at the end of the construction. We focus on two architectures: *(a)* shared-memory and shared-disk, which represents the modern multi-core desktop systems; and *(b)* shared-nothing architecture, such as computer clusters or cloud computing environments.

We use two metrics common to high performance computing: *(a) Strong scalability* (also known as speed-up): shows the performance for constant input size and increasing number of processors; and *(b) Weak scalability* (also known as scale-up): shows the performance when the ratio of the input size to the number of processors remains constant (e.g., when the length of the string doubles, the number of processors also doubles).

**Shared-memory and -disk architecture.** The scalability of the shared-memory and shared-disk architecture suffers from the interference at the memory bus and the I/O system. Here, we investigate this interference. We used the same machine as the previous section (i.e., Linux with two quad-core Intel CPUs at 2.67GHz and 24GB RAM), but we employed 1 to 8 cores and used 16GB RAM. The memory is divided equally among cores (1 core with 16GB, or 2 cores with 8GB RAM each, etc.).

For the next experiment we turn off the disk seek optimization (Section 4.4); the reason will become evident in the next paragraph. Figure 12(a) illustrates the execution time of ERA-No Seek and WaveFront for the Human Genome (i.e., strong scalability) with 16GB RAM. With 4 cores ERA-No Seek indexes the entire Human Genome in 19 minutes. ERA-No Seek scales well up to 4 cores (4GB RAM per core). In fact ERA is at least 1.5 times faster than WaveFront for up to 4 cores. However, ERA does not scale well to 8 cores. We believe the reason is that each of the 8 cores accesses only 2GB RAM, meaning that a smaller part of the tree is processed in one iteration. Therefore, each core requests data more frequently, leading to bottlenecks due to interference. In contrast, we believe WaveFront scales better to 8 cores (although in absolute time is still worse than ERA), because the CPU cost of the algorithm is higher. Therefore, it requests less frequently data from the disk, causing less interference.

To confirm our justification we run the same experiment with the larger DNA dataset (i.e., 4GBps). Now each core has more workload. Figure 12(b) illustrates that ERA-No Seek scales better to 8 cores. The same graph also shows the performance of ERA with the disk seek optimization turned on. With few cores, ERA-With Seek performs better, because it skips the parts of the input string that do not contain relevant input (see Section 4.4). For 8 cores, however, ERA-No Seek becomes better. This is due to the fact that each of the 8 cores work asynchronously on different

---
[8]http://www.mcs.anl.gov/research/projects/mpi

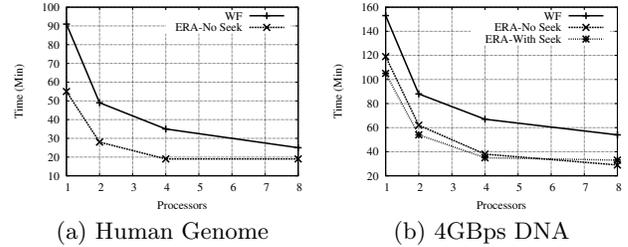

(a) Human Genome  (b) 4GBps DNA

**Figure 12:** Shared-memory, shared-disk; strong scalability; 16*GB* RAM divided equally among cores

| CPU | WaveFront (min) | ERA (min) | Gain | ERA speedup | ERA all speedup |
|---|---|---|---|---|---|
| 1 | 285.2 | 93.4 | 305% | – | – |
| 2 | 142.2 | 46.4 | 306% | 1 | 0.94 |
| 4 | 71.2 | 23.4 | 304% | 0.99 | 0.87 |
| 8 | 36.2 | 12.4 | 292% | 0.94 | 0.73 |
| 16 | 19.2 | 7.4 | 259% | 0.78 | 0.53 |

**Table 3:** Shared-nothing; strong scalability; human genome; 1GB RAM per CPU. The last column shows the speedup for the entire process. The other columns exclude the string transfer (2.3min) and the vertical partitioning phase (1.6min)

parts of the string. The disk seek optimization is applied independently by each core and causes the disk head to swing back and forth, creating significant delays.

**Shared-nothing architecture.** For the shared-nothing architecture experiments, we used a Linux cluster consisting of 16 machines connected through a switch, each with one dual-core Intel CPU at 3.33GHz and 8GB RAM. In each machine we used only one core and limited the memory to 1GB; therefore, if all 16 nodes are used, the total memory is 16GB. Note that the results in this paragraph are not comparable with those in the Share-memory and disk paragraph for two reasons: *(a)* the cluster machines have faster individual CPUs and *(b)* the total memory of the cluster varies from 1GB to 16GB depending on the number of machines used, whereas in the previous section the memory was fixed to 16GB irrespectively of the number of cores.

Table 3 shows the strong scalability results for the Human Genome. ERA is 3 times faster than WaveFront. Also the speed-up (refer to column titled ERA speedup) is very close to the theoretical optimal (i.e., 1.0), indicating very good load balancing. Note that all, but the last, columns in the table show only the time for tree construction. In the measurements, we have not included: *(i)* The time for the initial transfer of the input string to all nodes (roughly 2.3min). The bottleneck of string transfer is the slow switch; the performance can be improved with better network equipment that supports broadcast. *(b)* The vertical partitioning phase that takes around 1.6min since that phase has not been parallelized. The last column (titled ERA-all) shows the speedup considering these overheads; the speed-up is still very good, although not as close to the optimal. If everything is added, ERA indexes the entire Human Genome in roughly 11.3 minutes on a cluster with 16 commodity machines.

Since more memory was available in our machines, we run an experiment using 4GB per node. ERA indexed the Human Genome in 8.3 minutes. To the best of our knowledge in absolute time this is the fastest measurement reported so far. Note that we achieved this speed on a cluster whose total cost is roughly US$ 20,000 (February, 2011), low enough to be within the reach of many individuals.

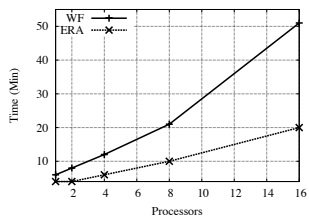

**Figure 13: Shared-nothing; weak scalability; DNA dataset; size varies from 256MBps to 4096MBps**

The last experiment investigates the weak scalability (recall that in weak scalability the ratio between the size of the input $S$ and the number of nodes is constant). We used the DNA dataset and varied the size from 256MBps for 1 node to 4096MBps for 16 nodes, with 1GB memory per node. According to the definition of weak scalability, in the optimal case the construction time should remain constant. In our case, since the size of $S$ increases proportionally to the number of nodes, the number of sub-trees to be constructed by each node is indeed constant. However, the average construction time of each sub-tree increases proportionally to $|S|$, because each node still needs to scan $S$ the same number of times. Therefore, optimal weak scalability cannot be achieved. Figure 13 illustrates that the construction time indeed increases linearly to the number of processors for ERA and WaveFront (note that the overheads for the string transfer is excluded). However, the rate of increase of ERA is much smaller than that of WaveFront. Therefore, as the string size increases, the performance gap widens; for example, when the string size is 4096MBps, ERA is 2.5 times faster than WaveFront. This is an important advantage of ERA, since in practice strings are expected to be very long.

## 7. CONCLUSIONS

Suffix trees are essential for many practical applications that include bioinformatics, processing of financial data (e.g., time series of stock market data), document clustering, etc. The volume of such data increases rapidly; therefore it is essential to have fast suffix tree construction methods. In this paper we proposed ERA, a method that supports very long strings, large alphabets, works efficiently even if memory is very limited and is easily parallelizable. Extensive experimental evaluation with very large real datasets revealed that our method is much more efficient than existing ones in terms of speed and computational resources. ERA indexes the entire human genome in 19 minutes on an ordinary 8-core desktop computer with 16GB RAM; and in 8.3min on a low-end cluster with 16 commodity computers with 4GB RAM each. To the best of our knowledge the fastest existing method (i.e., PWaveFront) needs 15min on an IBM BlueGene/L supercomputer with 1024 CPUs and 512GB RAM.

This work is part of a large project that aims to develop an engine for storing and processing of massive strings. We are currently working on scaling our method to thousands of CPUs. We are also focusing on the parallel processing of various types of queries using the suffix tree.